\documentclass[a4paper,11pt]{article}

\usepackage{graphicx,xcolor,amsmath,amssymb,amsthm,eufrak,babel,physics,fullpage}
\usepackage[hidelinks,colorlinks=true,linkcolor=blue,citecolor=blue,urlcolor=blue]{hyperref}
\usepackage{orcidlink}    
\usepackage[numbers,sort&compress]{natbib}
\newcommand{\eq}[2]{\begin{gather} #2  \label{#1} \end{gather}}
\newcommand{\nn}{\nonumber}
\newcommand{\Dt}{\mathcal{D}(t)}

\title{\bf On Prats' problem with anomalous diffusion}

\author{\bf A. Barletta%\footnote{corresponding author}
~\orcidlink{0000-0002-6994-5585}
\\[4pt]
{\small Department of Industrial Engineering, Alma Mater Studiorum Universit\`a di Bologna,} \\
{\small Viale Risorgimento 2, 40136 Bologna, Italy}\\ 
\small{\href{mailto:antonio.barletta@unibo.it}{antonio.barletta@unibo.it}}
}
\date{} % Activate to display a given date or no date (if empty),

\begin{document}
\maketitle

\begin{abstract}\noindent
The classical Prats' problem of flow instability in a horizontal porous channel saturated by a fluid subject to a buoyancy force is reconsidered. In the original formulation, the driving buoyancy force results from thermal diffusion. This study, however, substitutes thermal diffusion with mass diffusion. Furthermore, the usual scheme of mass diffusion is extended to comprehend also the anomalous phenomena of superdiffusion or subdiffusion. Such phenomena are modelled via a time-dependent mass diffusivity which yields a significant change in the formulation of the stability eigenvalue problem. In particular, the ordinary differential equations governing the time evolution of the perturbations acting on the base throughflow become non-autonomous. This makes a significant difference in the discussion of the conditions leading to instability, with a marked effect of the anomaly in the mass diffusion process. The transition from convective to absolute instability for subdiffusion processes is also addressed.\\[3mm]
\end{abstract}
\vspace{0.5cm}

%\begin{keywords}
%Authors should not enter keywords on the manuscript, as these must be chosen by the author during the online submission process and will then be added during the typesetting process (see \href{https://www.cambridge.org/core/journals/journal-of-fluid-mechanics/information/list-of-keywords}{Keyword PDF} for the full list).  Other classifications will be added at the same time.
%\end{keywords}

%{\bf MSC Codes }  76S05; 76R50; 76R10; 37N10

\section{Introduction}
The mechanism of mass diffusion in a fluid binary mixture is described on a microscopic scale as a molecular random walk (see, for example, \citealt{nelson1967dynamical}). The classical approach underlying the so-called normal mass diffusion is based on a very special relation between the variance, $\sigma_{\vb{x}}^2$, of the solute molecular position, $\vb{x}(t)$, and time, $t$. In fact, normal mass diffusion means a proportionality relation at large times,
\eq{int1}{
\sigma_{\vb{x}}^2 \sim \mathcal{D}\, t ,
}
where $\mathcal{D}$ is the mass diffusivity coefficient: a time-independent property of the binary mixture. Anomalous, or fractional, mass diffusion occurs when the behaviour of the random walk at large times means a departure from the proportionality law \eqref{int1}. Several studies and reviews have been published in the last decades regarding the physics and the mathematical modelling of anomalous diffusion \citep{metzler2000random, gorenflo2002discrete, gorenflo2002time, wu2008propagators, henry2010introduction, DOSSANTOS201986, cai2025machine}. The two ways where the large-time behaviour described by \eqref{int1} may become anomalous is through a faster spreading process of the molecular distribution, {\em i.e.}, superdiffusion, or through a slower spreading process, {\em i.e.}, subdiffusion. Thus, a usual model of anomalous diffusion is based on a power law where \eqref{int1} is actually replaced by 
\eq{int2}{
\sigma_{\vb{X}}^2 \sim \Dt\, t  \qc \qfor \Dt \sim t^{r-1}.
}
Here, $r$ is a positive index of anomalous diffusion. The special case $r=1$ means normal diffusion, with \eqref{int2} becoming equivalent to \eqref{int1}, while $r > 1$ defines superdiffusion and $0 < r < 1$ defines subdiffusion. As pointed out, for example, in \citet{BARLETTA2025109410} and in \citet{BarReCeBr2025}, the power law is just a special case of what the large time behaviour of the diffusivity can be for either subdiffusion or superdiffusion, whereas a more general definition will be adopted later on, in Section~\ref{linins} of this paper, with equation \eqref{19}. 
We refer the reader to the previous literature on the topic for specific assessments of the molecular dynamics that explain the reasons for the emergence of superdiffusive or subdiffusive processes. 

The aim of this paper is to make another step forward in the understanding of the interplay between anomalous mass diffusion and the emergence of buoyancy-induced flow instability and convection in a fluid saturated porous medium. This important subject was first investigated by \citet{karani2017onset} and by \citet{klimenko2017effect}. In those papers, anomalous diffusion is modelled via fractional differential operators. The authors either adopted a fractional-order advective term \citep{karani2017onset} in the diffusion equation or they subdivided the solute concentration into mobile and immobile phases linked by a kinetic equation involving a fractional time derivative \citep{klimenko2017effect}. Novel investigations of the Horton-Rogers-Lapwood problem \citep{horton1945convection, lapwood1948convection}, namely the onset of the Rayleigh-B\'enard instability in a fluid saturated horizontal porous layer, were recently proposed by \citet{barletta2023rayleigh} and by \citet{straughan2024asymptotic}. Up to this point, the analysis of the buoyancy-induced instability has only been developed for an equilibrium state of the fluid system where the seepage velocity is zero. The variant formulation of the Horton-Rogers-Lapwood problem where a uniform horizontal throughflow occurs at equilibrium is investigated in the pioneering paper by \citet{prats1966effect} and, after that study, it is now well-known as Prats' problem \citep{cheng1979heat, NiBe17}. 

The objective of the present study is to extend the analysis of Prats' problem, classically carried out for normal diffusion, with conditions where anomalous diffusion occurs. The P\'eclet number, which parametrises the throughflow rate, introduces an oscillatory pattern to the temporal evolution of the perturbations. The concurrent evolution of their amplitude then may lead the flow toward either stability or instability. Furthermore, the onset of instability is mainly influenced by the anomaly, superdiffusion or subdiffusion, and by the Rayleigh number for the cases of normal diffusion or subdiffusion. The possibility to detect the growing amplitude of the unstable perturbations in a laboratory reference frame is also discussed by investigating the transition from convective to absolute instability with special reference to the case of subdiffusion.

\section{Mathematical model}
Let us consider a horizontal porous channel saturated by a fluid binary mixture with $C$ the concentration of the solute, $\vb{u}$ the fluid seepage velocity and $P$ the local difference between the pressure and the hydrostatic pressure. The Cartesian components of $\vb{u}$ are $(u,v,w)$ along the $(x,y,z)$ axes. The buoyancy force acts on the fluid as described through the Oberbeck-Boussinesq approximation \citep{straughan2008stability, NiBe17, barletta2022boussinesq}, where the local momentum balance equation is written according to Darcy's law,
%passerini2014benard, 
%
%, arnone2025well
%
\eq{1}{
\div{\vb{u}} = 0,\nn\\
\frac{\mu}{K} \vb{u} = - \grad{P} - \rho_0\, \beta\, \qty(C - C_0)\, \vb{g},\nn\\
\varphi\pdv{C}{t} + \vb{u} \vdot \grad{C} = \varphi \,\Dt \laplacian{C}.
}
In \eqref{1}, the anomalous diffusion is modelled via a time-dependent mass diffusivity, $\Dt$ \citep{henry2010introduction, liang2021non, BARLETTA2025109410, BarReCeBr2025, girelli2025dynamical}. The properties $\mu$, $K$, $\beta$, $\varphi$ and $\rho_0$ are the dynamic viscosity, permeability, expansion coefficient relative to mass diffusion, porosity and fluid density evaluated when the solute concentration has the reference value $C_0$. The gravity acceleration $\vb{g}$ is uniform and parallel to the $z$ axis, so that $\vb{g} = - g\, \vb{e}_z$, with $g$ the modulus of $\vb{g}$ and $\vb{e}_z$ the unit vector of the $z$ axis.

The horizontal porous channel is confined by the horizontal impermeable planes, $z=0$ and $z=L$, kept at constant uniform concentrations $C_1$ and $C_2$, respectively. Thus, a reasonable choice of the reference concentration $C_0$ is the arithmetic average $\qty(C_1 + C_2)/2$. Let $\mathcal{D}_0$ be a conveniently chosen reference value of the time-dependent mass diffusivity. In the special case of normal diffusion, we choose the reference value $\mathcal{D}_0$ as the limit of $\Dt$ when $t \to +\infty$.
The governing equations \eqref{1} and their boundary conditions can be rendered dimensionless through the assignment of suitable scaling constants, namely
\eq{2}{
\frac{\qty(x,y,z)}{L} \to \qty(x ,  y ,  z) \qc 
\frac{t}{L^2/ \mathcal{D}_0} \to t \qc
 \frac{\vb{u}}{\varphi \, \mathcal{D}_0/L} \to \vb{u} , \nn\\
\frac{P}{\varphi \, \mu \mathcal{D}_0/K} \to P \qc \frac{C - C_0}{C_1 - C_2} \to C \qc \frac{\Dt}{\mathcal{D}_0} \to \Dt.
}
%
%where $t_0$ is an arbitrarily chosen instant of time. In the special case of normal diffusion, one may choose the reference value $\mathfrak{D}(t_0)$ as the limit of $\Dt$ when $t \to +\infty$.
%A convenient choice of $t_0$ we will adopt in the following is such that $t_0 = {L^2}/{\mathfrak{D}(t_0)}$.
We can now reformulate the governing equations in dimensionless terms as
\eq{4}{
\div{\vb{u}} = 0,\nn\\
\vb{u} = - \grad{P}  + Ra\, C\,  \vb{e}_z ,\nn\\
\pdv{C}{t} + \vb{u} \vdot \grad{C} = \Dt \laplacian{C}.
}
The mass diffusion Rayleigh number employed in \eqref{4} is defined as
\eq{5}{
Ra = \frac{\rho_0\, \beta\, g \qty(C_1 - C_2)\, K\, L}{\varphi \, \mu\, \mathcal{D}_0} . 
}
The dimensionless boundary conditions can be expressed as
\eq{6}{
z=0 : \qquad w = 0 \qc C = \frac{1}{2}, \nn\\
z=1 : \qquad w = 0 \qc C = - \frac{1}{2}.
}

\section{The stationary throughflow}
The Prats' problem is a variant of the Rayleigh-B\'enard problem for a saturated porous medium where the basic state features a steady horizontal throughflow. In fact, a basic solution of \eqref{4} and \eqref{6}, denoted with the subscript $b$, is given by
\eq{7}{
\vb{u}_b = \qty(Pe, 0, 0) \qc C_b = \frac{1}{2} - z \qc \grad{P}_b = \qty( - Pe, 0, Ra\, C_b ). 
}
The basic solution \eqref{7} is parametrised by the P\'eclet number, $Pe$, %a dimensionless parameter 
expressing the prescribed horizontal flow rate.

\section{Linear convective instability}\label{linins}
The linear convective instability of the basic throughflow \eqref{7} can be tested by superposing small-amplitude perturbations defined as
\eq{8}{
\vb{u} = \vb{u}_b + \epsilon\, \hat{\vb{u}} \qc C = C_b + \epsilon\, \hat{C} \qc \grad{P} = \grad{P}_b + \epsilon\, \grad{\hat{P}} \qc \qfor |\epsilon| \ll 1.
}
With \eqref{8}, by neglecting terms $\order{\epsilon^2}$, we can rewrite \eqref{4} and \eqref{6} as
\eq{9}{
\div{\hat{\vb{u}}} = 0,\nn\\
\hat{\vb{u}} = - \grad{\hat{P}}  + Ra\, \hat{C}\,  \vb{e}_z ,\nn\\
\pdv{\hat{C}}{t} - {\hat{w}} + Pe \, \pdv{\hat{C}}{x} = \Dt \laplacian{\hat{C}},
}
with
\eq{10}{
z=0, 1 : \qquad \hat{w} = 0 \qc \hat{C} = 0.
}
A pressure-concentration reformulation of \eqref{9} and \eqref{10} is obtained by evaluating the divergence of the second equation \eqref{9}, namely
\eq{11}{
\laplacian{\hat{P}}  - Ra\, \pdv{\hat{C}}{z} = 0,\nn\\
\pdv{\hat{C}}{t} + \pdv{\hat{P}}{z} - Ra\, \hat{C} + Pe \, \pdv{\hat{C}}{x} = \Dt \laplacian{\hat{C}},
}
with
\eq{12}{
z=0, 1 : \qquad \pdv{\hat{P}}{z} = 0 \qc \hat{C} = 0.
}
The strategy to solve the stability problem expressed by \eqref{11} and \eqref{12} is based on the use of normal modes,
\eq{13}{
\hat{P} = f(t) \cos(n \pi z) e^{i \qty(k_x x\, +\, k_y y)} \qc \hat{C} = h(t) \sin(n \pi z) e^{i \qty(k_x x\, +\, k_y y)},
%\qc k_x \in \mathbb{R} \qc k_y \in \mathbb{R}.
}
where $n$ is a positive integer, $\vb{k} =\qty(k_x, k_y, 0)$ is the horizontal wave vector and $k$ is the wave number, such that $k^2 = k_x^2 + k_y^2$. Equation \eqref{13} satisfies the boundary conditions \eqref{12} for every $n$ and $\vb{k}$, while equations \eqref{11} are satisfied if
\eq{14}{
f(t) = - \frac{n \pi\, Ra}{n^2\pi^2 + k^2} \, h(t) ,
}
and 
\eq{15}{
\dv{}{t} \log\!\qty[h(t)] = \frac{k^2 Ra}{n^2\pi^2 + k^2} - \qty(n^2\pi^2 + k^2)\, \Dt - i\, k_x\, Pe .
}
One can easily obtain the solution of \eqref{15},
\eq{16}{
h(t) = h(0)\, e^{\qty(n^2\pi^2 + k^2)\, G(t)}\, e^{-i\, k_x Pe\, t},
}
where
\eq{17}{
G(t) = \frac{k^2 Ra}{\qty(n^2\pi^2 + k^2)^2}\, t - \int\limits_0^t \mathcal{D}(t')\, \dd t' . 
}
What determines the ultimate effect of the perturbations superposed to the basic flow is the limit of $G(t)$ when $t \to +\infty$. In fact, we have three possible cases
\eq{18}{
\lim_{t \to +\infty} G(t) =
\begin{cases}
+ \infty \quad &\text{instability},\\
0  &\text{neutral stability},\\
- \infty  &\text{stability}.
\end{cases} 
}
Which of the three cases actually occurs depends strongly on the type of diffusion 
\eq{19}{
\lim_{t \to +\infty} \frac{1}{t} \int\limits_0^{t} \mathcal{D}(t')\, \dd t' =
\begin{cases}
+ \infty \qquad &\text{superdiffusion},\\
1  &\text{normal diffusion},\\
0  &\text{subdiffusion},
\end{cases} 
}
where, for normal diffusion, the value $1$ is a consequence of \eqref{2} and of the declared choice of $\mathcal{D}_0$.
On account of \eqref{17}-\eqref{19}, one may conclude that neutral stability is only possible for subdiffusion with $Ra=0$, or for normal diffusion provided that
\eq{20}{
Ra = \frac{\qty(n^2\pi^2 + k^2)^2}{k^2} .
}
Stability occurs either for superdiffusion independently of the Rayleigh number, for subdiffusion with $Ra < 0$, or for normal diffusion with
\eq{21}{
Ra < \frac{\qty(\pi^2 + k^2)^2}{k^2} .
}
Finally, instability takes place with subdiffusion for every positive Rayleigh number, or with normal diffusion when
\eq{22}{
Ra > \frac{\qty(\pi^2 + k^2)^2}{k^2} .
}
Equation \eqref{14} shows that the eigenfunctions, $f(t)$ and $h(t)$, are determined so that the initial values $f(0)$ and $h(0)$ must be linked in a special way. On the other hand, \eqref{16} forces the modulus of the eigenfunctions,  to evolve in time independently of the P\'eclet number and, hence, of the throughflow rate in the basic state \eqref{7}. As a consequence, the stable/unstable nature of the basic state is exactly the same as in the analysis of the anomalous Horton-Rogers-Lapwood problem carried out by \citet{barletta2023rayleigh}. The only difference is in the phase factor $\exp\qty(- i\, k_x Pe\, t)$ displayed in \eqref{16}. This term influences the ultimate asymptotic growth or decay of the perturbation amplitudes with an oscillatory behaviour. In practice, such oscillations are quite important, as they may hinder the possibility to detect, with a laboratory experiment, the actual growth of the perturbation amplitude with a measurement taken at a given spatial position in the streamwise direction $x$. This argument places the P\'eclet number on a central role in establishing experimentally the onset of instability. This aspect of the Prats' problem, in the case of normal diffusion, is pointed out in \citet{delache2007spatio} and in the book by \citet{barletta2019routes}. The effect of oscillations is grounded on the value of $k_x Pe$, so that this effect is maximum for transverse waves $(k_x = k, k_y = 0)$, while it gradually decreases for oblique waves and, eventually, vanishes with longitudinal waves $(k_x = 0, k_y = k)$. This well-known feature of flow instability results in the distinction between convective and absolute instability \citep{carriere1999convective, suslov2006numerical, delache2007spatio, barletta2019routes}. As explained in the literature relative to normal diffusion, absolute instability is possible for supercritical conditions, {\em i.e.}, when we are above the neutral stability threshold and $Ra$ exceeds a value, $Ra_{abs}$, that monotonically increases with $Pe$. Only when absolute instability occurs, an observer in a reference frame at rest is able to detect experimentally a perturbation amplifying in place and, hence, an unstable behaviour of the basic flow. In the forthcoming section, an extension of the methodology underlying the absolute instability analysis for processes of subdiffusion is presented in order to establish whether the predicted convective instability for $Ra > 0$ implies also absolute instability or not.

\section{Absolute instability for subdiffusion}
We follow the method described in Chapter~8 of the book by \citet{barletta2019routes}, and consider a wavepacket perturbation given by a superposition of normal modes defined by \eqref{13}, \eqref{14} and \eqref{16}. We consider a given horizontal direction of the normal modes, defined by an angle $\phi$ such that 
\eq{23}{
\vb{k} = \qty(k \cos\phi, k \sin\phi, 0) .
}
One may reckon from \eqref{17} and \eqref{19}, that, for subdiffusion, a large time approximation of $G(t)$ is 
\eq{24}{
G(t) \approx \frac{k^2 Ra}{\qty(n^2\pi^2 + k^2)^2}\, t .
}
Then, a wavepacket propagating along such a direction is described by the Fourier integral
\eq{25}{
\hat{C}(x,y,z,t) = \frac{1}{\sqrt{2\pi}} \int\limits_{-\infty}^{+\infty} A_{n,k}(x,y,z) \, e^{\lambda(k) t}\, \dd k ,
}
where
\eq{26}{
A_{n,k}(x,y,z) = h(0) \, \sin(n \pi z)\, e^{i k \qty(x \cos\phi\, +\, y \sin\phi)} ,
}
and
\eq{27}{
\lambda(k) = \frac{k^2 Ra}{n^2\pi^2 + k^2} - i\, k\, Pe \cos\phi .
}
It can be mentioned that a similar wavepacket expression applies also to $\hat{P}(x,y,z,t)$ with just two differences in the definition of $A_{n,k}(x,y,z)$: $f(0)$ must be used instead of $h(0)$ and $\cos(n \pi z)$ must be used instead of $\sin(n \pi z)$. 
We can now develop the steepest descent approximation for the wavepacket $\hat{C}(x,y,z,t)$ having in mind that the same outcome applies also to $\hat{P}(x,y,z,t)$ as it is based just on the coefficient $\lambda(k)$ which is common to both wavepackets. We have to determine the complex zeros, $k_0$, of the derivative $\dd \lambda(k)/\dd k$, {\em i.e.}, the saddle points. Thus, the saddle points, $k = k_0$, can be expressed in terms of the complex roots, $\gamma$, of
%%
%\eq{28}{
%\frac{2k\, n^2\pi^2 \, R}{\qty(n^2\pi^2 + k^2)^2} = i \qc \qfor R = \frac{Ra}{Pe \cos\phi} . 
%}
%%
%
\eq{28}{
\frac{2\, \gamma \, X}{\qty(1 + \gamma^2)^2} = i \qc \qfor \gamma =\frac{k}{n \pi} \qc X = \frac{Ra}{n \pi\, Pe \cos\phi} . 
}
By solving \eqref{28}, one can determine the saddle points for every given value of $X$. If we aim to determine the threshold $X = X_{abs}$ to absolute instability, then the solution of \eqref{28} must be sought so that $\Re(\lambda(k_0)) = 0$, where $\Re$ and $\Im$ stand for the real and imaginary parts of a complex variable. Such an additional constraint allows one to determine both the saddle points $k_0$ and the value of $X_{abs}$, namely
\eq{29}{
%k_0/(n\pi) = \pm 0.833783006304 - 1.04191281056\, i  \qc X_{abs} = 1.27032468110 , 
%\nn \\
%k_0/(n\pi) = \pm 0.833783006304 + 1.04191281056\, i  \qc X_{abs} = - 1.27032468110 .
k_0/(n\pi) = \pm 0.8337830 - 1.041913\, i  \qc X_{abs} = 1.270325 , 
\nn \\
k_0/(n\pi) = \pm 0.8337830 + 1.041913\, i  \qc X_{abs} = - 1.270325 .
}
%
%The lowest threshold of $|X_{abs}|$ is attained with the modes $n=1$, so that we can just consider this case. 
%%
%\eq{30}{
%k_0 = \qty(\pm 0.833783006304 - 1.04191281056\, i) \pi  \qc R_{abs} = 1.27032468110 \, \pi, 
%\nn \\
%k_0 = \qty(\pm 0.833783006304 + 1.04191281056\, i) \pi  \qc R_{abs} = - 1.27032468110\, \pi .
%}
%%
Given that the threshold value of the Rayleigh number $Ra$ for the onset of absolute instability cannot be negative, $Ra_{abs} \ge 0$, while there are no restrictions on the sign of $Pe\,\cos\phi$, both the symmetric values of $X_{abs}$ given by \eqref{29} are meaningful. We can conclude that absolute instability for subdiffusion occurs when $Ra$ exceeds a threshold, $Ra_{abs}$, that can be expressed as
\eq{30}{
Ra_{abs} = 1.270325 \, n \pi\, |Pe \cos\phi|.
}
Then, the least Rayleigh number conditions for absolute instability corresponding to a prescribed P\'eclet number are achieved with $n=1$ and $\cos\phi = 0$, namely with longitudinal modes, $\phi = \pi/2$ or $\vb{k} = \qty(0, k, 0)$. On the other hand, the largest threshold $Ra_{abs}$ to absolute instability corresponds to transverse modes, $\phi = 0$ or $\vb{k} = \qty(k, 0, 0)$. This result makes perfect sense as longitudinal modes grow in place so that they are not convected downstream by the basic throughflow. Stated differently, the time evolution of the longitudinal modes is unaffected by the oscillatory phase factor, $\exp\qty(- i\, k_x Pe\, t)$, displayed in \eqref{16}.

\begin{figure}[t]
\centering
\includegraphics[width=0.4\textwidth]{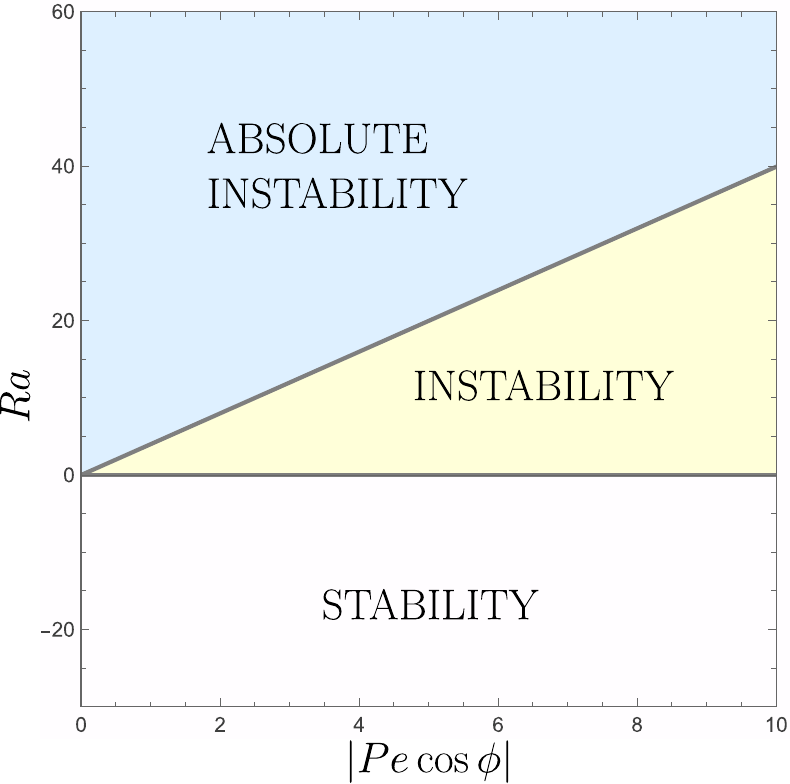}
\caption{\label{fig1}Subdiffusion: results of the linear analysis of instability}
\end{figure}

\begin{figure}[ht]
\centering
\includegraphics[width=0.6\textwidth]{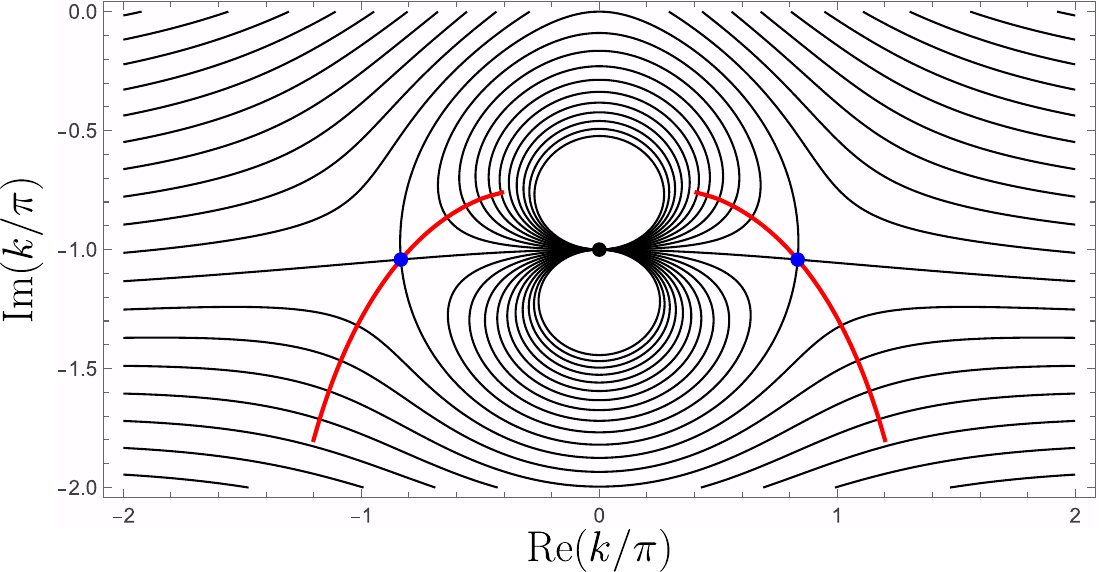}
\caption{\label{fig2}Check of the holomorphy requirement: isolines of $\Re\qty(\lambda(k))$ in the complex $k$ plane, for $Ra = Ra_{abs}$ with $n=1$ and $Pe \cos\phi > 0$. The blue dots correspond to the saddle points $k=k_0$, while the black dot denotes the singularity $k = -i \pi$. The red lines show the paths of steepest descent crossing the saddle points}
\end{figure}
%\begin{figure}[h!]
%\centering
%\includegraphics[width=0.75\textwidth]{fig2b}
%\caption{\label{fig2b}Check of the holomorphy requirement: sketches of the original integration path (green line) and of a sample deformed path including the lines of steepest descent (blue and red lines) and crossing the saddle points}
%\end{figure}

\begin{figure}[h!]
\centering
\includegraphics[width=0.4\textwidth]{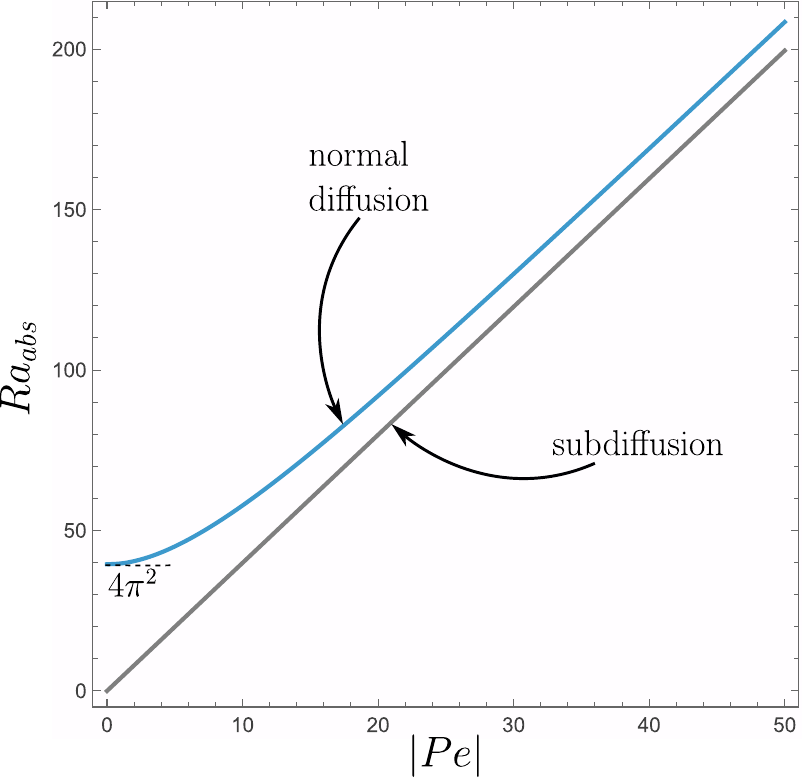}
\caption{\label{fig3}Transverse modes $(\phi = 0)$ with $n=1$: comparison between the absolute instability threshold for normal diffusion \citep{delache2007spatio, barletta2019routes} and that for subdiffusion}
\end{figure}

Figure~\ref{fig1} illustrates the results of the linear stability analysis for subdiffusion as they emerge from the arguments presented in Section~\ref{linins} and from the evaluation of the absolute instability threshold given by equation \eqref{30} with $n=1$. 

Figures~\ref{fig2} %and \ref{fig2b} 
shows graphically that the so-called holomorphy requirement \citep{suslov2006numerical, barletta2019routes} is satisfied by the saddle points given by \eqref{29}. This means that one can continuously deform the integration domain for the right hand side of \eqref{25}, namely the line $\Im(k) = 0$, into a curve crossing the saddle points through a path which is locally of steepest descent for $\Re\qty(\lambda(k))$. Such paths of steepest descent are displayed in Fig.~\ref{fig2} %and \ref{fig2b} 
as red lines. The blue dots are the saddle points, while the black dot is the singularity $k = -i \pi$. The saddle points are at the intersections of the isolines $\Re\qty(\lambda(k)) = 0$. The union of the deformed path crossing the saddle points and the original integration path, $\Im(k) = 0$, must yield a contour that does not contain any singularity of $\Re\qty(\lambda(k))$. Figure~\ref{fig2} demonstrates that the latter statement holds true or, equivalently, that the holomorphy requirement can be satisfied. 

Figure~\ref{fig3} displays a comparison between the trends of $Ra_{abs}$ for the case of transverse modes with normal diffusion (data reported by \citet{delache2007spatio} and \citet{barletta2019routes}) with equation~\eqref{30} for subdiffusion. One can conclude that the transition to absolute instability for normal diffusion requires larger values of $Ra$ than for subdiffusion, for every fixed value of $Pe$. As suggested by Fig.~\ref{fig3}, for large values of $Pe$, the difference, $\Delta Ra_{abs}$, between $Ra_{abs}$ for normal diffusion and subdiffusion tends to an asymptotic value that can be evaluated by employing Richardson's extrapolation, namely $\Delta Ra_{abs} \approx 7.3356$. A common feature of normal diffusion and subdiffusion displayed in Fig.~\ref{fig3} is that, with $Pe=0$, the transition to absolute instability takes place at critical conditions without any gap in $Ra$, {\em i.e.}, with $Ra = Ra_c = 4\pi^2$ for normal diffusion and $Ra = Ra_c = 0$ for subdiffusion.

\section{Conclusions}
The anomalous diffusion effects on the onset of instability in a horizontal porous channel flow have been investigated. The mechanism of instability relies on the interplay between mass diffusion and buoyancy-force as modelled through the Oberbeck-Boussinesq approximation. The model adopted is based on Darcy's law of momentum transfer and on a mass diffusion equation accounting both for advection and for the anomalous diffusion phenomenon, through a time-dependent diffusivity coefficient. The instability for the equilibrium solution of the local balance governing equations has been tested. The equilibrium solution is a basic state of stationary flow with a uniform seepage velocity in a horizontal direction and a uniform vertical concentration gradient. Perturbations superposed to such an equilibrium state have been defined as normal modes with a given dimensionless wave vector along any horizontal direction. The transition to instability is governed by the Rayleigh number, $Ra$, and the P\'eclet number, $Pe$. 
The main findings can be summarised as follows:
\begin{itemize}
\item Superdiffusion, modelled as a process where the mass diffusivity averaged over a large time diverges to infinity, leads to a decay in time of the perturbation amplitudes, for every real value of $Ra$ and for every real value of $Pe$, and hence to stability.
\item Subdiffusion, modelled as a process where the mass diffusivity averaged over a large time tends to zero, leads to an amplification in time of the perturbation amplitudes, for every positive $Ra$ and for every real value of $Pe$, and hence to instability. With $Ra=0$, subdiffusion leads to neutral stability and, with $Ra < 0$, to stability.
\item For normal diffusion, modelled as a process where the mass diffusivity averaged over a large time tends to a nonzero constant, we recover exactly the same results described in the paper by \citet{prats1966effect} and reviewed in \citet{cheng1979heat} and \citet{NiBe17}.
\item The nonzero throughflow envisaged in the basic equilibrium state, parametrised through the P\'eclet number, yields an oscillatory trend to the amplification or decay of the perturbation amplitude for either transverse or oblique modes. 
%The frequency of such oscillations, proportional to $Pe$, is at its greatest for wavelike perturbations propagating along the streamwise horizontal direction (transverse normal modes), while it is zero for wavelike perturbations propagating along the horizontal direction perpendicular to the basic throughflow (longitudinal normal modes).
\item Transition to absolute instability for subdiffusion occurs for $Ra > Ra_{abs}$ with $Ra_{abs}$ directly proportional to $|Pe|$. For every fixed $Pe$, the transition to absolute instability occurs at lower Rayleigh numbers for subdiffusion than for normal diffusion.
\end{itemize}
There are several opportunities for future developments of the research on anomalous diffusion and the instability of flows. In fact, the analysis of the anomalous Prats' problem is a mathematically simple application, while more complicated situations may emerge for the Navier-Stokes shear flows with mass diffusion. For instance, the emergence of an interplay between the hydrodynamic instability and the anomalous diffusion phenomenology may display novel features for the so-called Rayleigh-B\'enard-Poiseuille problem.

\section*{Acknowledgements}
This work was supported by Alma Mater Studiorum Universit\`a di Bologna, grant number RFO-2025.

%\bibliography{biblio}

\begin{thebibliography}{27}
\expandafter\ifx\csname natexlab\endcsname\relax\def\natexlab#1{#1}\fi
\providecommand{\bibinfo}[2]{#2}
\ifx\xfnm\relax \def\xfnm[#1]{\unskip,\space#1}\fi
%Type = Book
\bibitem[{Nelson(1967)}]{nelson1967dynamical}
\bibinfo{author}{E.~Nelson}, \bibinfo{title}{Dynamical Theories of Brownian
  Motion}, \bibinfo{publisher}{Princeton University Press},
  \bibinfo{address}{Princeton, NJ}, \bibinfo{year}{1967}.
%Type = Article
\bibitem[{Metzler and Klafter(2000)}]{metzler2000random}
\bibinfo{author}{R.~Metzler}, \bibinfo{author}{J.~Klafter},
\newblock \bibinfo{title}{The random walk's guide to anomalous diffusion: a
  fractional dynamics approach},
\newblock \bibinfo{journal}{Physics Reports} \bibinfo{volume}{339}
  (\bibinfo{year}{2000}) \bibinfo{pages}{1--77}.
%Type = Article
\bibitem[{Gorenflo et~al.(2002{\natexlab{a}})Gorenflo, Mainardi, Moretti,
  Pagnini, and Paradisi}]{gorenflo2002discrete}
\bibinfo{author}{R.~Gorenflo}, \bibinfo{author}{F.~Mainardi},
  \bibinfo{author}{D.~Moretti}, \bibinfo{author}{G.~Pagnini},
  \bibinfo{author}{P.~Paradisi},
\newblock \bibinfo{title}{Discrete random walk models for space-time fractional
  diffusion},
\newblock \bibinfo{journal}{Chemical Physics} \bibinfo{volume}{284}
  (\bibinfo{year}{2002}{\natexlab{a}}) \bibinfo{pages}{521--541}.
%Type = Article
\bibitem[{Gorenflo et~al.(2002{\natexlab{b}})Gorenflo, Mainardi, Moretti, and
  Paradisi}]{gorenflo2002time}
\bibinfo{author}{R.~Gorenflo}, \bibinfo{author}{F.~Mainardi},
  \bibinfo{author}{D.~Moretti}, \bibinfo{author}{P.~Paradisi},
\newblock \bibinfo{title}{Time fractional diffusion: a discrete random walk
  approach},
\newblock \bibinfo{journal}{Nonlinear Dynamics} \bibinfo{volume}{29}
  (\bibinfo{year}{2002}{\natexlab{b}}) \bibinfo{pages}{129--143}.
%Type = Article
\bibitem[{Wu and Berland(2008)}]{wu2008propagators}
\bibinfo{author}{J.~Wu}, \bibinfo{author}{K.~M. Berland},
\newblock \bibinfo{title}{Propagators and time-dependent diffusion coefficients
  for anomalous diffusion},
\newblock \bibinfo{journal}{Biophysical Journal} \bibinfo{volume}{95}
  (\bibinfo{year}{2008}) \bibinfo{pages}{2049--2052}.
%Type = Incollection
\bibitem[{Henry et~al.(2010)Henry, Langlands, and
  Straka}]{henry2010introduction}
\bibinfo{author}{B.~I. Henry}, \bibinfo{author}{T.~A.~M. Langlands},
  \bibinfo{author}{P.~Straka},
\newblock \bibinfo{title}{An introduction to fractional diffusion},
\newblock in: \bibinfo{editor}{R.~L. Dewar}, \bibinfo{editor}{F.~Detering}
  (Eds.), \bibinfo{booktitle}{Complex Physical, Biophysical and Econophysical
  Systems}, \bibinfo{publisher}{World Scientific},
  \bibinfo{address}{Singapore}, \bibinfo{year}{2010}, pp.
  \bibinfo{pages}{37--89}.
%Type = Article
\bibitem[{{dos Santos}(2019)}]{DOSSANTOS201986}
\bibinfo{author}{M.~A.~F. {dos Santos}},
\newblock \bibinfo{title}{Analytic approaches of the anomalous diffusion: a
  review},
\newblock \bibinfo{journal}{Chaos, Solitons \& Fractals} \bibinfo{volume}{124}
  (\bibinfo{year}{2019}) \bibinfo{pages}{86--96}.
%Type = Article
\bibitem[{Cai et~al.(2025)Cai, Hu, Qu, Zhao, Wang, Li, and
  Huang}]{cai2025machine}
\bibinfo{author}{W.~Cai}, \bibinfo{author}{Y.~Hu}, \bibinfo{author}{X.~Qu},
  \bibinfo{author}{H.~Zhao}, \bibinfo{author}{G.~Wang},
  \bibinfo{author}{J.~Li}, \bibinfo{author}{Z.~Huang},
\newblock \bibinfo{title}{Machine learning analysis of anomalous diffusion},
\newblock \bibinfo{journal}{The European Physical Journal Plus}
  \bibinfo{volume}{140} (\bibinfo{year}{2025}) \bibinfo{pages}{183}.
%Type = Article
\bibitem[{Barletta et~al.(2025{\natexlab{a}})Barletta, Brandão, Capone, and
  {De Luca}}]{BARLETTA2025109410}
\bibinfo{author}{A.~Barletta}, \bibinfo{author}{P.~V. Brandão},
  \bibinfo{author}{F.~Capone}, \bibinfo{author}{R.~{De Luca}},
\newblock \bibinfo{title}{Stabilizing effect of generic anomalous diffusion
  independent of the {R}ayleigh number},
\newblock \bibinfo{journal}{International Communications in Heat and Mass
  Transfer} \bibinfo{volume}{168} (\bibinfo{year}{2025}{\natexlab{a}})
  \bibinfo{pages}{109410}.
%Type = Article
\bibitem[{Barletta et~al.(2025{\natexlab{b}})Barletta, Rees, Celli, and
  {Brand\~ao}}]{BarReCeBr2025}
\bibinfo{author}{A.~Barletta}, \bibinfo{author}{D.~A.~S. Rees},
  \bibinfo{author}{M.~Celli}, \bibinfo{author}{P.~V. {Brand\~ao}},
\newblock \bibinfo{title}{Open boundaries, anomalous diffusion and the
  {D}arcy‐{B\'e}nard instability},
\newblock \bibinfo{journal}{Transport in Porous Media} \bibinfo{volume}{152}
  (\bibinfo{year}{2025}{\natexlab{b}}) \bibinfo{pages}{27}.
%Type = Article
\bibitem[{Karani et~al.(2017)Karani, Rashtbehesht, Huber, and
  Magin}]{karani2017onset}
\bibinfo{author}{H.~Karani}, \bibinfo{author}{M.~Rashtbehesht},
  \bibinfo{author}{C.~Huber}, \bibinfo{author}{R.~L. Magin},
\newblock \bibinfo{title}{Onset of fractional-order thermal convection in
  porous media},
\newblock \bibinfo{journal}{Physical Review E} \bibinfo{volume}{96}
  (\bibinfo{year}{2017}) \bibinfo{pages}{063105}.
%Type = Article
\bibitem[{Klimenko and Maryshev(2017)}]{klimenko2017effect}
\bibinfo{author}{L.~S. Klimenko}, \bibinfo{author}{B.~S. Maryshev},
\newblock \bibinfo{title}{Effect of solute immobilization on the stability
  problem within the fractional model in the solute analog of the
  {H}orton-{R}ogers-{L}apwood problem},
\newblock \bibinfo{journal}{The European Physical Journal E}
  \bibinfo{volume}{40} (\bibinfo{year}{2017}) \bibinfo{pages}{1--7}.
%Type = Article
\bibitem[{Horton and Rogers~Jr(1945)}]{horton1945convection}
\bibinfo{author}{C.~W. Horton}, \bibinfo{author}{F.~T. Rogers~Jr},
\newblock \bibinfo{title}{Convection currents in a porous medium},
\newblock \bibinfo{journal}{Journal of Applied Physics} \bibinfo{volume}{16}
  (\bibinfo{year}{1945}) \bibinfo{pages}{367--370}.
%Type = Article
\bibitem[{Lapwood(1948)}]{lapwood1948convection}
\bibinfo{author}{E.~R. Lapwood},
\newblock \bibinfo{title}{Convection of a fluid in a porous medium},
\newblock \bibinfo{journal}{Mathematical Proceedings of the Cambridge
  Philosophical Society} \bibinfo{volume}{44} (\bibinfo{year}{1948})
  \bibinfo{pages}{508--521}.
%Type = Article
\bibitem[{Barletta(2023)}]{barletta2023rayleigh}
\bibinfo{author}{A.~Barletta},
\newblock \bibinfo{title}{Rayleigh-{B\'e}nard instability in a horizontal
  porous layer with anomalous diffusion},
\newblock \bibinfo{journal}{Physics of Fluids} \bibinfo{volume}{35}
  (\bibinfo{year}{2023}) \bibinfo{pages}{104114}.
%Type = Article
\bibitem[{Straughan and Barletta(2024)}]{straughan2024asymptotic}
\bibinfo{author}{B.~Straughan}, \bibinfo{author}{A.~Barletta},
\newblock \bibinfo{title}{Asymptotic behaviour for convection with anomalous
  diffusion},
\newblock \bibinfo{journal}{Continuum Mechanics and Thermodynamics}
  \bibinfo{volume}{36} (\bibinfo{year}{2024}) \bibinfo{pages}{737--743}.
%Type = Article
\bibitem[{Prats(1966)}]{prats1966effect}
\bibinfo{author}{M.~Prats},
\newblock \bibinfo{title}{The effect of horizontal fluid flow on thermally
  induced convection currents in porous mediums},
\newblock \bibinfo{journal}{Journal of Geophysical Research}
  \bibinfo{volume}{71} (\bibinfo{year}{1966}) \bibinfo{pages}{4835--4838}.
%Type = Incollection
\bibitem[{Cheng(1978)}]{cheng1979heat}
\bibinfo{author}{P.~Cheng},
\newblock \bibinfo{title}{Heat transfer in geothermal systems},
\newblock volume~\bibinfo{volume}{14} of \textit{\bibinfo{series}{Advances in
  Heat Transfer}}, \bibinfo{publisher}{Academic Press}, \bibinfo{address}{New
  York, NY}, \bibinfo{year}{1978}, pp. \bibinfo{pages}{1--105}.
%Type = Book
\bibitem[{Nield and Bejan(2017)}]{NiBe17}
\bibinfo{author}{D.~A. Nield}, \bibinfo{author}{A.~Bejan},
  \bibinfo{title}{Convection in Porous Media}, \bibinfo{publisher}{Springer},
  \bibinfo{address}{New York, NY}, \bibinfo{edition}{5th} edition,
  \bibinfo{year}{2017}.
%Type = Book
\bibitem[{Straughan(2008)}]{straughan2008stability}
\bibinfo{author}{B.~Straughan}, \bibinfo{title}{Stability and Wave Motion in
  Porous Media}, \bibinfo{publisher}{Springer}, \bibinfo{address}{New York,
  NY}, \bibinfo{year}{2008}.
%Type = Article
\bibitem[{Barletta(2022)}]{barletta2022boussinesq}
\bibinfo{author}{A.~Barletta},
\newblock \bibinfo{title}{The {B}oussinesq approximation for buoyant flows},
\newblock \bibinfo{journal}{Mechanics Research Communications}
  \bibinfo{volume}{124} (\bibinfo{year}{2022}) \bibinfo{pages}{103939}.
%Type = Article
\bibitem[{Liang et~al.(2021)Liang, Tian, Wang, and Xu}]{liang2021non}
\bibinfo{author}{Y.~Liang}, \bibinfo{author}{P.~Tian},
  \bibinfo{author}{S.~Wang}, \bibinfo{author}{W.~Xu},
\newblock \bibinfo{title}{Non-{F}ickian diffusion in time-space fluctuating
  diffusivity landscapes: From superfast to ultraslow},
\newblock \bibinfo{journal}{Fractals} \bibinfo{volume}{29}
  (\bibinfo{year}{2021}) \bibinfo{pages}{2150191}.
%Type = Article
\bibitem[{Girelli et~al.(2025)Girelli, Giantesio, Musesti, and
  Penta}]{girelli2025dynamical}
\bibinfo{author}{A.~Girelli}, \bibinfo{author}{G.~Giantesio},
  \bibinfo{author}{A.~Musesti}, \bibinfo{author}{R.~Penta},
\newblock \bibinfo{title}{Dynamical anomalous transport of molecules subject to
  inhomogeneous body forces},
\newblock \bibinfo{journal}{Zeitschrift f{\"u}r Angewandte Mathematik und
  Physik} \bibinfo{volume}{76} (\bibinfo{year}{2025}) \bibinfo{pages}{249}.
%Type = Article
\bibitem[{Delache et~al.(2007)Delache, Ouarzazi, and
  Combarnous}]{delache2007spatio}
\bibinfo{author}{A.~Delache}, \bibinfo{author}{M.~N. Ouarzazi},
  \bibinfo{author}{M.~Combarnous},
\newblock \bibinfo{title}{Spatio-temporal stability analysis of mixed
  convection flows in porous media heated from below: Comparison with
  experiments},
\newblock \bibinfo{journal}{International Journal of Heat and Mass Transfer}
  \bibinfo{volume}{50} (\bibinfo{year}{2007}) \bibinfo{pages}{1485--1499}.
%Type = Book
\bibitem[{Barletta(2019)}]{barletta2019routes}
\bibinfo{author}{A.~Barletta}, \bibinfo{title}{Routes to Absolute Instability
  in Porous Media}, \bibinfo{publisher}{Springer}, \bibinfo{address}{New York,
  NY}, \bibinfo{year}{2019}.
%Type = Article
\bibitem[{Carriere and Monkewitz(1999)}]{carriere1999convective}
\bibinfo{author}{P.~Carriere}, \bibinfo{author}{P.~A. Monkewitz},
\newblock \bibinfo{title}{Convective versus absolute instability in mixed
  {R}ayleigh-{B\'e}nard-{P}oiseuille convection},
\newblock \bibinfo{journal}{Journal of Fluid Mechanics} \bibinfo{volume}{384}
  (\bibinfo{year}{1999}) \bibinfo{pages}{243--262}.
%Type = Article
\bibitem[{Suslov(2006)}]{suslov2006numerical}
\bibinfo{author}{S.~A. Suslov},
\newblock \bibinfo{title}{Numerical aspects of searching convective/absolute
  instability transition},
\newblock \bibinfo{journal}{Journal of Computational Physics}
  \bibinfo{volume}{212} (\bibinfo{year}{2006}) \bibinfo{pages}{188--217}.

\end{thebibliography}

\end{document}